\documentstyle[twoside,fleqn,espcrc2]{article}

\newcommand{\be}{\begin{equation}}
\newcommand{\ee}{\end{equation}}
\newcommand{\ba}{\begin{eqnarray}}
\newcommand{\ea}{\end{eqnarray}}
\newcommand{\Tr}{{\rm Tr}}
\newcommand{\Dmrns}{{\cal D}_{\mu\rho,\nu\sigma}}

\newcommand{\AmS}{{\protect\the\textfont2
  A\kern-.1667em\lower.5ex\hbox{M}\kern-.125emS}}

\title{Field strength correlators in QCD at zero and 
 non--zero temperature}

\author{A. Di Giacomo$^{\,\, \rm a}$,
        E. Meggiolaro \address{Dipartimento di Fisica
        dell'Universit\`a and I.N.F.N., Sezione di Pisa, \\ 
        I--56100 Pisa, Italy.}
        \thanks{Speaker at the conference.}
        and
        H. Panagopoulos \address{Department of Natural Sciences,
        University of Cyprus, \\
        1678 Nicosia, Cyprus.}
        }

\begin{document}

\begin{abstract}
We study, by numerical simulations on a lattice, the behaviour of the 
gauge--invariant field strength correlators in QCD both at zero temperature, 
down to a distance of 0.1 fm, and at finite temperature, across the 
deconfinement phase transition.
\end{abstract}

\maketitle

\section{INTRODUCTION}

An important role in hadron physics is played by
the gauge--invariant two--point correlators of the field strengths in the 
QCD vacuum. They are defined as
\be
\Dmrns(x) = \langle 0| 
\Tr \left\{ G_{\mu\rho}(x) S G_{\nu\sigma}(0) S^\dagger \right\}
|0\rangle ~,
\ee
where $G_{\mu\rho} = gT^aG^a_{\mu\rho}$ is the field--stength tensor
and $S = S(x,0)$ is the Schwinger phase operator needed to 
parallel--transport the tensor $G_{\nu\sigma}(0)$ to the point $x$.

These correlators govern the effect of the gluon condensate on the level 
splittings in the spectrum of heavy $Q \bar{Q}$ bound states
\cite{Gromes82,Campostrini86,Simonov95}.
They are the basic quantities in models of stochastic confinement of colour
\cite{Dosch87,Dosch88,Simonov89}
and in the description of high--energy hadron scattering
\cite{Nachtmann84,Landshoff87,Kramer90,Dosch94}.

A numerical determination of the correlators on lattice (with gauge group
$SU(3)$) already exists, in the range of physical distances between 0.4 and 1 
fm \cite{DiGiacomo92}. In that range $\Dmrns$ falls off exponentially 
\be
\Dmrns(x) \sim \exp(-|x|/\lambda) ~,
\ee
with a correlation length $\lambda \simeq 0.22$ fm \cite{DiGiacomo92}.

What makes the determination of the correlators possible on the lattice,
with a reasonable computing power, is the idea \cite{Campostrini89,DiGiacomo90}
of removing the effects of short--range fluctuations on 
large distance correlators by a local {\it cooling} procedure.
Freezing the links of QCD configurations one after the other, damps very 
rapidly the modes of short wavelength, but requires a number $n$ of cooling 
steps proportional to the square of the distance $d$ in lattice units to 
affect modes of wavelength $d$:
\be
n \simeq k d^2 ~.
\ee
Cooling is a kind of diffusion process.
If $d$ is sufficiently large, there will be a range of values of $n$ in 
which lattice artefacts due to short--range fluctuations have been removed, 
without touching the physics at distance $d$; by {\it lattice artefacts} we 
mean statistical fluctuations and renormalization effects from lattice to 
continuum. This removal will show up as a plateau in the dependence 
of the correlators on $n$. This was the technique successfully used in
Ref. \cite{DiGiacomo92}. There, the range of distances explored was from 
from 3--4 up to 7--8 lattice spacings at $\beta \simeq 6$ ($\beta = 6/g^2$), 
which means approximately from 0.4 up to 1 fm in physical distance.
The lattice size was $16^4$.

In Sect. 2 we discuss new results obtained on a $32^4$ lattice, at $\beta$
between 6.6 and 7.2: at these values of $\beta$ the lattice size is still 
bigger than 1 fm, and therefore safe from infrared artefacts, but
$d = 3,4$ lattice spacings now correspond to physical distances of about
0.1 fm. Since what matters to our cooling procedure is the distance in 
lattice units, we obtain in this way a determination of the correlators at 
distances down to 0.1 fm \cite{myref1}.

All of the above concerns the theory at zero temperature.
In Sect. 3 we go further and determine the behaviour of the correlators at 
finite temperature for the pure--gauge theory with $SU(3)$ colour group and in 
particular we study their behaviour across the deconfining phase transition
\cite{myref2}. The motivations to do that stem from Refs. 
\cite{Simonov1,Simonov2,Simonov3}.

\section{RESULTS AT $T = 0$}

The most general form of the correlator compatible with the invariances of 
the system at zero temperature is \cite{Dosch87,Dosch88,Simonov89}
\ba
\lefteqn{
\Dmrns(x) = (\delta_{\mu\nu}\delta_{\rho\sigma} - \delta_{\mu\sigma}
\delta_{\rho\nu}) \left[ {\cal D}(x^2) + {\cal D}_1(x^2) \right] } \nonumber \\
& & + (x_\mu x_\nu \delta_{\rho\sigma} - x_\mu x_\sigma \delta_{\rho\nu} 
+ x_\rho x_\sigma \delta_{\mu\nu} \nonumber \\
& & - x_\rho x_\nu \delta_{\mu\sigma})
{\partial{\cal D}_1(x^2) \over \partial x^2} ~.
\ea
${\cal D}$ and ${\cal D}_1$ are invariant functions of $x^2$. We work in 
the Euclidean region.

It is convenient to define a ${\cal D}_\parallel(x^2)$ and a 
${\cal D}_\perp(x^2)$ as follows:
\ba
{\cal D}_\parallel &\equiv& {\cal D} + {\cal D}_1 + x^2 {\partial{\cal D}_1
\over \partial x^2} ~, \nonumber \\
{\cal D}_\perp &\equiv& {\cal D} + {\cal D}_1 ~.
\ea
On the lattice we can define a lattice operator $\Dmrns^L$, which is
proportional to $\Dmrns$ in the na\"\i ve continuum limit, i.e., when the 
lattice spacing $a \to 0$.
Making use of the definition (5) we can thus write, in the same limit,
\ba
{\cal D}_\parallel^L(\hat d a) \mathop\sim_{a\to0} a^4 
{\cal D}_\parallel(d^2 a^2) + {\cal O}(a^6) ~,\nonumber \\
{\cal D}_\perp^L(\hat d a) \mathop\sim_{a\to0} a^4 
{\cal D}_\perp(d^2 a^2) + {\cal O}(a^6) ~.
\ea
Equations (6)
come from a formal expansion of the operator, and are expected to be modified, 
when the expectation value is computed, by lattice artefacts, i.e., by 
effects due to the ultraviolet {\it cutoff}. These effects can be estimated in 
perturbation theory and subtracted \cite{Campostrini84}. 
Instead we remove them by cooling the quantum fluctuations at the scale of 
the lattice spacing, as explained in the Introduction. 
After cooling, ${\cal D}_\parallel^L$ and ${\cal D}_\perp^L$ are expected to 
obey Eqs. (6).
The typical behaviour of ${\cal D}_\parallel^L$ and of ${\cal D}_\perp^L$ 
along cooling is shown in Fig. 1.
Our data are the values of the correlations at the plateau.

\begin{figure}[t]
\vspace{4.5cm}
\includegraphics{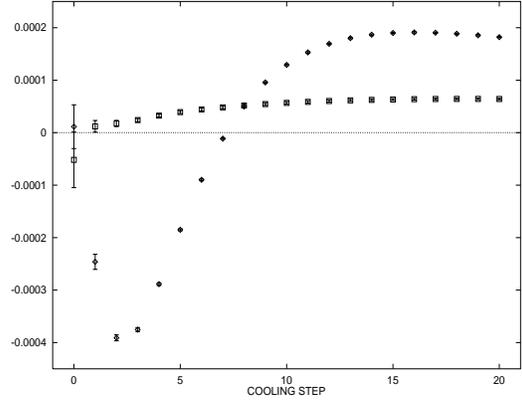} 
\null\vskip 0.3cm
\caption{A typical behaviour of ${\cal D}_\parallel^L$
(diamonds; $d = 6$, $\beta = 6.6$, lattice $32^4$)
and of ${\cal D}_\perp^L$ (squares; $d = 12$, $\beta = 6.6$, lattice $32^4$)
during cooling.}
\end{figure}

\begin{figure}[t]
\vspace{4.5cm}
\includegraphics{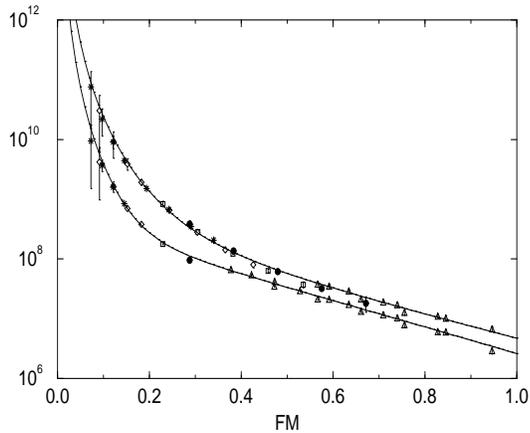} 
\null\vskip 0.3cm
\caption{The functions ${\cal D}_\perp^L f(\beta)^{-4}$ (upper curve) and 
${\cal D}_\parallel^L f(\beta)^{-4}$ (lower curve)
versus physical distance (in {\it fermi} units). Triangles correspond to the 
data of Ref. [11]. The lines are the curves for 
${\cal D}_\perp$ and ${\cal D}_\parallel$ obtained from the best fit of Eqs. 
(10) and (11).}
\end{figure}

We have measured the correlations on a $32^4$ lattice at distances ranging 
from 3 to 14 lattice spacings and at $\beta = 6.6,~6.8,~7.0,~7.2$. 
From renormalization group arguments,
\be
a(\beta) = {1\over\Lambda_L} f(\beta) ~,
\ee
where $\Lambda_L$ is the fundamental mass--scale of QCD in the lattice 
renormalization scheme.
At large enough $\beta$, $f(\beta)$ is given by the usual two--loop 
expression:
\be
f(\beta) \simeq \left({8\over33}\,\pi^2\beta\right)
^{ 51/121 } \exp\left(-{4\over33}\pi^2\beta\right) ~,
\ee
for gauge group $SU(3)$ and in the absence of quarks.
At sufficiently large $\beta$ one also expects that
\ba
{\cal D}_\parallel^L f(\beta)^{-4} &=& {1\over\Lambda_L^4}
{\cal D}_\parallel\left({d^2\over\Lambda_L^2}f^2(\beta)\right) ~,
\nonumber \\
{\cal D}_\perp^L f(\beta)^{-4} &=& {1\over\Lambda_L^4}
{\cal D}_\perp\left({d^2\over\Lambda_L^2}f^2(\beta)\right) ~,
\ea
where $f(\beta)$ is given by Eq. (8) and terms of higher order in $a$ are 
negligible.

In Fig. 2 we plot ${\cal D}_\parallel^L f(\beta)^{-4}$ 
and ${\cal D}_\perp^L f(\beta)^{-4}$ versus $d_{\rm phys} = 
(d/\Lambda_L)$  $f(\beta)$. In this figure we have also plotted the values 
of the correlators obtained in Ref. \cite{DiGiacomo92}, corresponding to 
physical distances $d_{\rm phys} \ge 0.4$ fm.
We have applied a best fit to all of these data with the functions
\ba
\lefteqn{
{\cal D}_{(1)} (x^2) = A_{(1)} \exp\left( -|x|/\lambda_A \right) }
\nonumber \\
& & + {a_{(1)} \over |x|^4} \exp\left( -|x|/\lambda_a \right) ~.
\ea
We have obtained the following results:
\ba
{A \over \Lambda_L^4} \simeq 3.3 \times 10^8 &~,~&
{A_1 \over \Lambda_L^4} \simeq 0.7 \times 10^8 ~, \nonumber \\
a \simeq 0.69 &~,~& a_1 \simeq 0.46 ~, \nonumber \\
\lambda_A \simeq {1\over\Lambda_L}{1\over182} &~,~&
\lambda_a \simeq {1\over\Lambda_L}{1\over94} ~,
\ea
with $\chi^2/N_{\rm d.o.f.} \simeq 1.7$. The continuum lines in Fig. 2
have been obtained using the parameters of this best fit.
With the value of $\Lambda_L$
determined from the string tension \cite{Michael88} we obtain
\be
\lambda_A \simeq 0.22 \, {\rm fm} ~,~ \lambda_a \simeq 0.43 \, {\rm fm} ~.
\ee
The correlation length $\lambda_A$, which enters the non--perturbative
exponential terms of ${\cal D}$ and ${\cal D}_1$, as well as the magnitude 
of the coefficients $A$ and $A_1$, are compatible with the values obtained
in Ref. \cite{DiGiacomo92}.

We stress again that we have been able to observe terms 
proportional to $1/|x|^4$ in the correlations because we have worked at 
larger values of $\beta$, where the distance between two points 
(far enough in lattice units so that the correlation is not modified by 
cooling before lattice artefacts are eliminated) is small compared with 
$1\,{\rm fm}$ in physical units. A larger lattice ($32^4$) has been 
necessary to avoid infrared artefacts.

\section{RESULTS AT FINITE $T$}

To simulate the system at finite temperature, a lattice is used of spatial 
extent $N_\sigma \gg N_\tau$, $N_\tau$ being the temporal extent, 
with periodic boundary conditions.
The temperature $T$ corresponding to a given value of $\beta = 6/g^2$ is 
given by
\be
N_\tau \cdot a(\beta) = {1 \over T} ~,
\ee
where $a(\beta)$ is the lattice spacing, whose expression in terms of 
$\beta$ is given by Eqs. (7) and (8).

At finite temperature, the $O(4)$ space--time symmetry is broken down to
the spatial $O(3)$ symmetry and in principle the bilocal correlators (1)
are now expressed in terms of five independent functions
\cite{Simonov1,Simonov2,Simonov3} (instead of two as in the 
zero--temperature case); two of them are needed to
describe the electric--electric correlations:
\ba
\lefteqn{
\langle 0| \Tr \left\{ E_i (x) S(x,y) E_k (y) S^\dagger(x,y) \right\}
|0\rangle = } \nonumber \\
& & \delta_{ik} \left[ D^E + D_1^E + u_4^2 {\partial D_1^E \over 
\partial u_4^2} \right] + u_i u_k {\partial D_1^E \over 
\partial \vec{u}^2} ,
\ea
where $E_i = G_{i4}$ is the electric field operator and 
$u_\mu = x_\mu - y_\mu$ [$\vec{u}^2 = (\vec{x} - \vec{y})^2$].

Two further functions are needed for the magnetic--magnetic correlations:
\ba
\lefteqn{
\langle 0| \Tr \left\{ B_i (x) S(x,y) B_k (y) S^\dagger(x,y) \right\}
|0\rangle = } \nonumber \\
& & \delta_{ik} \left[ D^B + D_1^B + \vec{u}^2 {\partial D_1^B \over 
\partial \vec{u}^2} \right] - u_i u_k {\partial D_1^B \over 
\partial \vec{u}^2} ,
\ea
where $B_k = {1 \over 2} \varepsilon_{ijk} G_{ij}$ is the magnetic field
operator. Finally, one more function, $D_1^{BE}$, is necessary to describe 
the mixed electric--magnetic correlations.

The five quantities $D^E$, $D_1^E$, $D^B$, 
$D_1^B$ and $D_1^{BE}$ are all functions of $\vec{u}^2$, due to rotational 
invariance, and of $u_4^2$, due to time--reversal invariance.

From the conclusions of Refs. \cite{Simonov1,Simonov2,Simonov3}, 
one expects that $D^E$ is related to the (temporal)
string tension and should have a drop just above the deconfinement critical 
temperature $T_c$. In other words, $D^E$ is expected to be the order 
parameter of the confinement.
Similarly, $D^B$ is related to the spatial string tension 
\cite{Simonov1,Simonov2}.

We have determined the following four quantities
\ba
{\cal D}_\parallel^E (\vec{u}^2,0) &\equiv&
 {\cal D}^E (\vec{u}^2,0) + {\cal D}_1^E (\vec{u}^2,0) + 
\vec{u}^2 {\partial{\cal D}_1^E \over \partial \vec{u}^2}
\nonumber \\
{\cal D}_\perp^E (\vec{u}^2,0) &\equiv& {\cal D}^E (\vec{u}^2,0)
+ {\cal D}_1^E (\vec{u}^2,0)
\nonumber \\
{\cal D}_\parallel^B (\vec{u}^2,0) &\equiv&
 {\cal D}^B (\vec{u}^2,0) + {\cal D}_1^B (\vec{u}^2,0) 
+ \vec{u}^2 {\partial{\cal D}_1^B \over \partial \vec{u}^2}
\nonumber \\
{\cal D}_\perp^B (\vec{u}^2,0) &\equiv& {\cal D}^B (\vec{u}^2,0) + 
{\cal D}_1^B (\vec{u}^2,0) ~,
\ea
by measuring appropriate linear superpositions of the correlators (14) 
and (15) at equal times ($u_4 = 0$), on a $16^3 \times 4$ lattice 
($N_\tau = 4$, in our notation). 
The critical temperature $T_c$ for such a lattice corresponds to
$\beta_c \simeq 5.69$.
The behaviour of ${\cal D}_\parallel^E$ and ${\cal D}_\perp^E$ is shown 
in Figs. 3 and 4 respectively, on three--dimensional plots versus
$T/T_c$ and the physical spatial distance.
Due to the logarithmic scale, the errors are comparable with
the size of the symbols and the lines connecting the points are drawn as an 
eye--guide. A clear drop is observed for ${\cal D}_\parallel^E$ and 
${\cal D}_\perp^E$ across the phase transition, as expected.

\begin{figure}[t]
\vspace{4.5cm}
\includegraphics{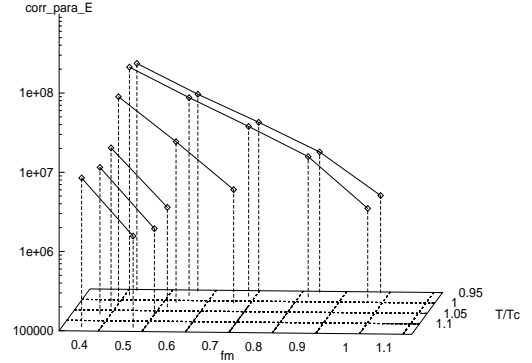} 
\null\vskip 0.3cm
\caption{The quantity ${\cal D}_\parallel^E / \Lambda_L^4$
[defined by the first Eq. (16)] versus $T/T_c$ and the physical spatial
distance (in fm).}
\end{figure}

\begin{figure}[t]
\vspace{4.5cm}
\includegraphics{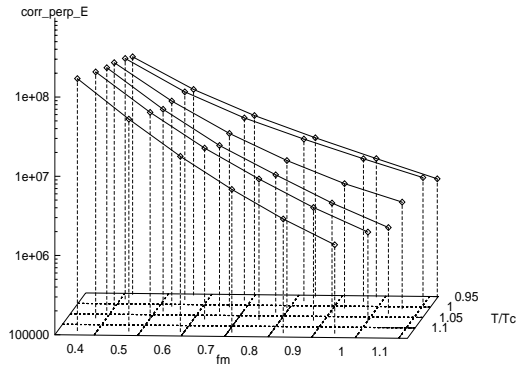} 
\null\vskip 0.3cm
\caption{The quantity ${\cal D}_\perp^E / \Lambda_L^4$
[defined by the second Eq. (16)] versus $T/T_c$ and the physical spatial
distance (in fm).}
\end{figure}

On the contrary, no drop is visible across the transition for 
the magnetic correlations ${\cal D}_\parallel^B$ and ${\cal D}_\perp^B$: 
as an example, the behaviour for ${\cal D}_\perp^B$ is shown in Fig. 5.


\begin{figure}[t]
\vspace{4.5cm}
\includegraphics{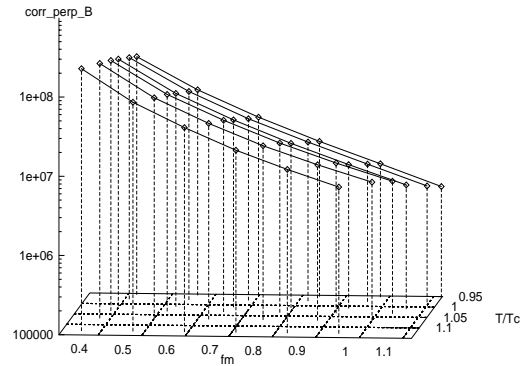} 
\null\vskip 0.3cm
\caption{The quantity ${\cal D}_\perp^B / \Lambda_L^4$
[defined by the fourth Eq. (16)] versus $T/T_c$ and the physical spatial
distance (in fm).}
\end{figure}

Our results can be summarized as follows:

(i) In the confined phase ($T < T_c$), until very near to the temperature of 
deconfinement, the correlators, both the electric--electric type (14) and
the magnetic--magnetic type (15), are nearly equal to the correlators at 
zero temperature \cite{myref1}: in other words, $D^E \simeq D^B
\simeq D$ and $D_1^E \simeq D_1^B \simeq D_1$ for $T < T_c$.

(ii) Immediately above $T_c$, the electric--electric correlators (14) have 
a clear drop, while the magnetic--magnetic correlators (15) stay almost 
unchanged, or show a slight increase. 
More precisely, looking at the values for the difference
\be
{\cal D}_\perp^E (\vec{u}^2,0) - {\cal D}_\parallel^E (\vec{u}^2,0) =
-\vec{u}^2 {\partial{\cal D}_1^E \over \partial \vec{u}^2} (\vec{u}^2,0)
\ee
between the two quantities reported in Figs. 4 and 3 respectively, one 
finds that the quantity ${\cal D}_1^E$ does not show any drop across the 
phase transition at $T_c$. So the clear drop seen in the quantities
${\cal D}_\parallel^E$ and ${\cal D}_\perp^E$ across $T_c$ is entirely due 
to a drop of ${\cal D}^E$ alone. This result confirms the conclusion of
Refs. \cite{Simonov1,Simonov2,Simonov3}, where ${\cal D}^E$ was related to 
the (temporal) string tension $\sigma_E$.
Similarly, one can look at the following difference:
\be
{\cal D}_\perp^B (\vec{u}^2,0) - {\cal D}_\parallel^B (\vec{u}^2,0) =
-\vec{u}^2 {\partial{\cal D}_1^B \over \partial \vec{u}^2} (\vec{u}^2,0) ~.
\ee
One thus 
finds that ${\cal D}_1^B$ does not show any drop across the transition and, 
in addition, it is nearly equal to ${\cal D}_1^E$ (${\cal D}_1^B \simeq
{\cal D}_1^E$). From the fact that the quantities ${\cal D}_\parallel^B$ 
and ${\cal D}_\perp^B$ stay almost unchanged (or even show a slight 
increase) across $T_c$, we conclude that the same must be true also for
${\cal D}^B$. It was shown in Refs. \cite{Simonov1,Simonov2} that 
${\cal D}^B$ is related to the spatial string tension $\sigma_s$. 
Recent lattice results \cite{Laermann95} indicate that $\sigma_s$ is 
almost constant around $T_c$ and increases for $T \ge 2 T_c$: this fact is in 
good agreement with the behaviour that we have found for ${\cal D}^B$.

\bigskip
\noindent {\bf ACKNOWLEDGEMENTS}
\smallskip

This work was done using the CRAY T3D of the CINECA Inter University 
Computing Centre (Bologna, Italy). We would like to thank the CINECA for 
having put the CRAY T3D at our disposal and for the kind and highly qualified 
technical assistance. 

We thank G\"unther Dosch and Yuri Simonov for many 
useful discussions.


\begin{thebibliography}{9}
\bibitem{Gromes82}
D. Gromes, Phys. Lett. {\bf 115B} (1982) 482.
\bibitem{Campostrini86}
M. Campostrini, A. Di Giacomo and S. Olejnik,
Z. Phys. {\bf C31} (1986) 577.
\bibitem{Simonov95}
Yu.A. Simonov, S. Titard and F.J. Yndurain, Phys. Lett. {\bf B354} 
(1995) 435.
\bibitem{Dosch87}
H.G. Dosch, Phys. Lett. {\bf 190B} (1987) 177.
\bibitem{Dosch88}
H.G. Dosch and Yu.A. Simonov, Phys. Lett. {\bf 205B} (1988) 339.
\bibitem{Simonov89}
Yu.A. Simonov, Nucl. Phys. {\bf B324} (1989) 67.
\bibitem{Nachtmann84}
O. Nachtmann and A. Reiter, Z. Phys. {\bf C24} (1984) 283.
\bibitem{Landshoff87}
P.V. Landshoff and O. Nachtmann, Z. Phys. {\bf C35} (1987) 405.
\bibitem{Kramer90}
A. Kr\"amer and H.G. Dosch, Phys. Lett. {\bf 252B} (1990) 669.
\bibitem{Dosch94}
H.G. Dosch, E. Ferreira and A. Kr{\"a}mer, Phys. Rev. D {\bf 50} (1994) 
1992.
\bibitem{DiGiacomo92}
A. Di Giacomo and H. Panagopoulos, Phys. Lett. {\bf B285} (1992) 133.
\bibitem{Campostrini89}
M. Campostrini, A. Di Giacomo, M. Maggiore, H. Panagopoulos and E. Vicari,
Phys. Lett. {\bf 225B} (1989) 403.
\bibitem{DiGiacomo90}
A. Di Giacomo, M. Maggiore and S. Olejnik,
Phys. Lett {\bf 236B} (1990) 199;
Nucl. Phys. {\bf B347} (1990) 441.
\bibitem{myref1}
A. Di Giacomo, E. Meggiolaro and H. Panagopoulos, Pisa preprint IFUP--TH 
12/96 (1996); Cyprus preprint UCY--PHY--96/5 (1996); hep--lat/9603017.
\bibitem{myref2}
A. Di Giacomo, E. Meggiolaro and H. Panagopoulos, Pisa preprint IFUP--TH 
14/96 (1996); Cyprus preprint UCY--PHY--96/6 (1996); hep--lat/9603018.
\bibitem{Simonov1}
Yu.A. Simonov, JETP Lett. {\bf 54} (1991) 249.
\bibitem{Simonov2}
Yu.A. Simonov, JETP Lett. {\bf 55} (1992) 627;
Yad. Fiz. {\bf 58} (1995) 357.
\bibitem{Simonov3}
Yu.A. Simonov and E.L. Gubankova, Phys. Lett. {\bf B360} (1995) 93.
\bibitem{Campostrini84}
M. Campostrini, A. Di Giacomo and G. Mussardo,
Z. Phys. {\bf C25} (1984) 173.
\bibitem{Michael88}
C. Michael and M. Teper, Nucl. Phys. {\bf B305} (1988) 453.
\bibitem{Laermann95}
E. Laermann, Nucl. Phys. B (Proc. Suppl.) {\bf 42} (1995) 120,
and references therein.
\end{thebibliography}
\end{document}